\documentclass[%
reprint,
superscriptaddress,
 amsmath,amssymb,
 aps,
 pra,
]{revtex4-2}

\usepackage{graphicx}
\usepackage{dcolumn}
\usepackage{bm}
\usepackage{comment}
\usepackage{makecell}

\begin{document}

\preprint{He**NIR/PRA26}

\title{Photoelectron spectroscopy of 3s3p doubly excited helium\\ dressed with strong near-infrared laser fields}%

\author{Mizuho Fushitani}%
\affiliation{%
Department of Chemistry, Graduate School of Science, Nagoya University, Nagoya, Aichi 464-8602, Japan
}%
\affiliation{%
RIKEN, SPring-8 Center, Sayo, Hyogo 679-5148, Japan
}%

\author{Chien-Nan Liu}
\affiliation{%
Department of Physics, Fu-Jen Catholic University, Taipei 24205, Taiwan
}%

\author{Yuki Ono}%
\affiliation{%
Department of Chemistry, Graduate School of Science, Nagoya University, Nagoya, Aichi 464-8602, Japan
}%

\author{Shunsuke Amaike}%
\affiliation{%
Department of Chemistry, Graduate School of Science, Nagoya University, Nagoya, Aichi 464-8602, Japan
}%

\author{Wataru Yamazaki}%
\affiliation{%
Department of Chemistry, Graduate School of Science, Nagoya University, Nagoya, Aichi 464-8602, Japan
}%

\author{Keiko Kato}%
\affiliation{%
Department of Chemistry, Graduate School of Science, Nagoya University, Nagoya, Aichi 464-8602, Japan
}%

\author{Akitaka Matsuda}%
\affiliation{%
Department of Chemistry, Graduate School of Science, Nagoya University, Nagoya, Aichi 464-8602, Japan
}%
\affiliation{%
RIKEN, SPring-8 Center, Sayo, Hyogo 679-5148, Japan
}%

\author{Shigeki Owada}%
\affiliation{%
RIKEN, SPring-8 Center, Sayo, Hyogo 679-5148, Japan
}%
\affiliation{%
Japan Synchrotron Radiation Research Institute, Sayo, Hyogo 679-5198, Japan
}%

\author{Makina Yabashi}%
\affiliation{%
RIKEN, SPring-8 Center, Sayo, Hyogo 679-5148, Japan
}%
\affiliation{%
Japan Synchrotron Radiation Research Institute, Sayo, Hyogo 679-5198, Japan
}%

\author{Yasumasa Hikosaka}
\affiliation{
Institute of Liberal Arts and Sciences, University of Toyama, Toyama 930-0194, Japan
}%
\affiliation{%
RIKEN, SPring-8 Center, Sayo, Hyogo 679-5148, Japan
}%

\author{Toru Morishita}
\affiliation{Institute for Advanced Science, The University of Electro-Communications, 1-5-1 Chofu-ga-oka, Chofu-shi, Tokyo 182-8585, Japan}%

\author{Akiyoshi Hishikawa}
\email{hishikawa.akiyoshi.z6@f.mail.nagoya-u.ac.jp}
\affiliation{%
Department of Chemistry, Graduate School of Science, Nagoya University, Nagoya, Aichi 464-8602, Japan
}%
\affiliation{%
RIKEN, SPring-8 Center, Sayo, Hyogo 679-5148, Japan
}%
\affiliation{%
Research Center for Materials Science, Nagoya University, Nagoya, Aichi 464-8602, Japan
}%

\date{\today}

\begin{abstract}
We report time-resolved photoelectron spectroscopy of the $3s3p$ doubly excited states of helium dressed by a strong near-infrared (NIR) laser field. 
Using synchronized XUV free-electron-laser and 800-nm NIR laser pulses, we observe a pronounced delay-dependent shift of resonance-related spectral minima together with the emergence of additional structures around the NIR sideband energy. 
\textit{Ab initio} theoretical calculations support these observations and identify the features as signatures of NIR-induced coupling of the bright ($3s3p {}^{1}P^{o}$) autoionizing state to nearby dark ($^{1}D^{e}$ and $^{1}S^{e}$) resonances below the $N = 3$ threshold. 
A multichannel Fano resonance analysis of the measured spectra yields delay-dependent line-shape parameters and resonance energies, establishing a quantitative route to characterize and control correlated two-electron resonances in strong laser fields.
\end{abstract}

\maketitle


\section{Introduction}
The helium atom, consisting of two electrons and a nucleus, serves as an ideal platform for studying a Coulombic three-body system. 
In particular, doubly excited states constitute as a benchmark to understand the electron-electron correlation \cite{Fano1961, Fano1983, Lin1995,Rost1997, Tanner2000} and its control by strong external laser fields \cite{Lambropoulos1998}.
The doubly excited states exhibit asymmetric resonance profiles arising from quantum interference between the decay pathways from a discrete quasi-bound state and the ionization continuum state. 
Below the $N=2$ threshold of He$^+$ with $N$ being the principal quantum number, the absorption cross section at photon energy $E$ is expressed as \cite{Fano1961}
\begin{equation}
    \sigma(E) = \sigma_0 \frac{(q+\varepsilon)^2}{1+\varepsilon^2},
\end{equation}
where $\varepsilon = 2(E-E_r)/\Gamma$ with $E_r$ and $\Gamma$ denoting the resonance energy and the autoionization width of the doubly excited state, respectively, and $q$ is the Fano line shape parameter. 
This expression yields one minimum and one maximum in the spectra (unless $q = 0$ or $\infty$) to produce an asymmetric resonance profile.

Doubly excited states of helium in strong laser fields have been investigated extensively over the last decades \cite{Loh2008, Gilbertson2010, Argenti2010, Chu2011, Chu2012, Ott2013, Chu2014, Ott2014, Kaldun2014, Argenti2015, Zielinski2015, Artemyev2017, Petersson2017, Ott2019, Argenti2023,Zhang2025,He2025}. 
Experimental and theoretical manipulation of spectral line shapes has been demonstrated by transient absorption spectroscopy \cite{Loh2008,Ott2013,Chu2012,He2025} as well as by photoelectron spectroscopy \cite{Chu2011,Chu2012}.
Laser-assisted resonances to dark states have also been discussed theoretically \cite{Chu2014, Argenti2015}.
Strong laser fields have been used to prepare coherent superposition of the doubly excited states \cite{Argenti2010, Kaldun2014, Ott2014} and to investigate their effect on the excited-state lifetimes \cite{Aufleger2020, Rupprecht2024}.

Most of these studies focus on the $2s2p$ $^1P^{o}$ state below the $N = 2$ threshold of He$^+$.
Doubly excited states conversing to the $N = 3$ threshold \cite{Starace1977, Zhou1994, Domke1996} offer an additional platform to explore the correlated electron dynamics and its control.
Ultrafast excitation of $N = 3$ excited states was demonstrated with an XUV free-electron laser (FEL) via three-photon absorption at $\hbar\omega_\mathrm{XUV}$ = 24.1 eV \cite{Hishikawa2011}.
The excitation proceeds by a one-photon resonant transition that promotes one electron to a Rydberg state, followed by two-photon excitation of the inner electron.
This provides a unique route to create the two-electron wavepackets, opening the possibility of visualizing the electron-electron correlation in the time domain \cite {Liu2012}.
More recently, a theoretical study discussed the coherent dynamics of $N = 3$ doubly excited-states coupled by a strong near-infrared (NIR) laser pulse at 798 nm, using attosecond transient absorption at $\hbar\omega_\mathrm{XUV}$ = 68 eV \cite{Petersson2017}.
It predicted clear beatings in the absorption spectrum as a function of the pump-probe time delay $\Delta t$ between the XUV and NIR laser pulses.
The dominant one-photon beatings involving the $3s3p$ and $3s4p$ $^1P^{o}$ states were found to occur with $^1S^{e}$ states. 
In addition, multichannel two-electron correlated wavepackets were suggested, which appear as nonresonant two-photon beatings between $^1P^{o}$ states.

\begin{figure}[tb]
\includegraphics[width=8.5cm]{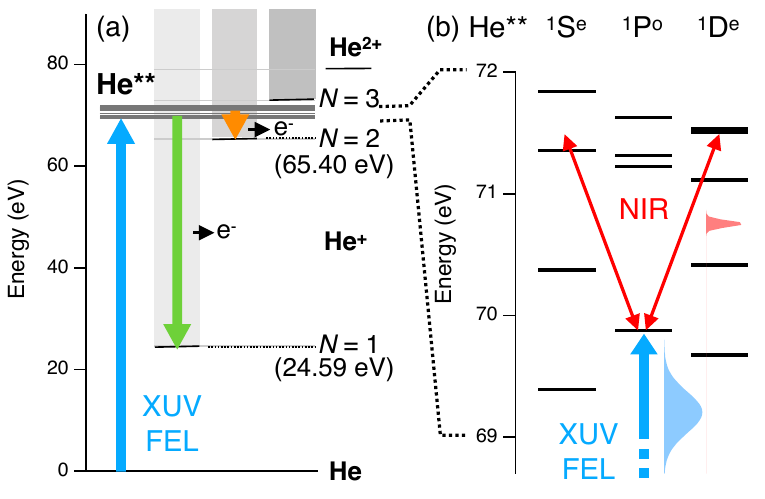}
\caption{
Energy diagram of electronic states of He relevant to the present study. (a) Overview and (b) expanded view of the doubly excited states. XUV-FEL energy is tuned to a $^1P^{o}$ state, which is coupled to $^1D^{e}$ and $^1S^{e}$ states by strong NIR laser fields (800 nm). 
}
\label{fig:energydiagram} 
\end{figure}

Here we present photoelectron spectroscopy of the $N = 3$ doubly excited states of He dressed by strong NIR laser fields ($\sim 10^{12}$ W/cm$^2$).
The energy diagram of the relevant states is shown in Fig.~\ref{fig:energydiagram}.
In this manifold, the bright $3s3p~^1P^{o}$ state populated by XUV can be coupled by a single NIR photon to neighboring dark states below the $N = 3$ threshold.
We identify resonances associated with $^1D^{e}$ and $^1S^{e}$ states and quantify the delay-dependent shift of the effective resonance energy in the photoelectron spectra, providing a channel-resolved view of NIR dressing in correlated two-electron dynamics.

\begin{figure*}[tb]
\includegraphics[width=15cm]{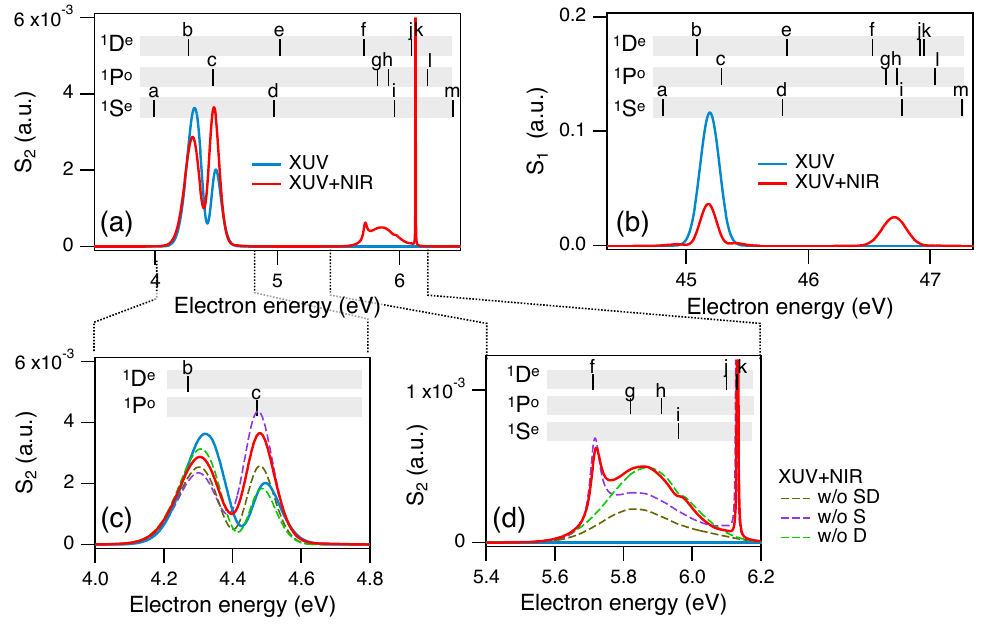}
\caption{
Theoretical photoelectron spectra of NIR-dressed doubly excited He, $S_{N}(E_e, \omega_\mathrm{XUV})$ with $\omega_\mathrm{XUV} = 2.564$~a.u., for decay into the He$^+$ (a)$N = 2$ and (b)$N = 1$ channels, calculated for XUV only and for XUV+NIR at zero delay $\Delta t = 0$~fs. 
Panels (c) and (d) show expanded views of the $N = 2$ spectra in the main-peak and sideband regions, respectively. 
Also shown are results of additional calculations with truncated bases, where the quasi-bound hyperspherical channels of $^{1}D^{e}$ (``w/o D”), $^{1}S^{e}$ (``w/o S”) or of both symmetries (``w/o SD”) converging to the He$^+$ $(N \ge 3)$ thresholds, are removed. 
They exhibit distinct differences, highlighting the role of these resonances. 
No pronounced resonance-related changes are observed in the $N = 1$ spectra.}\label{fig:theospectra} 
\end{figure*}

\begin{table}[t]
\centering
\caption{
Doubly excited states of helium converging to the $N = 3$ threshold of He$^+$, relevant to the present study.
The $_N(K, T)_n^\pm$ notation \cite{Lin1995} is also listed, where $N$ and $n$ denote the principal quantum numbers of the inner and outer electrons, respectively, and $(K, T)$ represent their angular correlation.
The superscript $\pm$ indicates the radial correlation of the two electrons, corresponding to in-phase ($+$) and out-of-phase ($-$) oscillations.
Energies relative to the $N = 2$ threshold of He$^+$ (in eV) are tabulated for the present hyperspherical calculation and Ref.~\cite{Ho1985}.\\
}
\label{tab:assignments}
\begin{tabular}{c c c c c c}
\hline
Label & Term & $_N(K,T)_n^\pm$ &  Energy &  \makecell{Energy\\ Ref.\cite{Ho1985}} \\
\hline
a & $^1S^{e}$ & $_3(2,0)_3^+$  &       & 3.99 \\
b & $^1D^{e}$ & $_3(2,0)_3^+$  &       & 4.27 \\
c & $^1P^{o}$ & $_3(1,1)_3^+$  & 4.47  & 4.47 \\
d & $^1S^{e}$ & $_3(0,0)_3^+$  &       & 4.97 \\
e & $^1D^{e}$ & $_3(0,2)_3^+$  &       & 5.02 \\
f & $^1D^{e}$ & $_3(0,0)_3^+$  & 5.72  & 5.71 \\
g & $^1P^{o}$ & $_3(2,0)_4^-$  &       & 5.82 \\
h & $^1P^{o}$ & $_3(-1,1)_3^+$ &       & 5.91 \\
i & $^1S^{e}$ & $_3(2,0)_4^+$  & 5.96  & 5.96 \\
j & $^1D^{e}$ & $_3(2,0)_4^+$  &       & 6.10 \\
k & $^1D^{e}$ & $_3(1,1)_4^-$  & 6.13  & 6.13 \\
l & $^1P^{o}$ & $_3(1,1)_4^+$  &       & 6.23 \\
m & $^1S^{e}$ & $_3(0,0)_4^+$  &       & 6.44 \\
\hline

\end{tabular}
\end{table}

\section{Experimental}
Photoelectron measurements were performed at the soft X-ray beamline (BL1) of the XFEL facility SACLA \cite{Yabashi2013}. 
Ultrashort XUV-FEL pulses ($\sim$70 eV, $\sim$30 fs \cite{Owada2020}) were delivered at a repetition rate of 60 Hz.
The XUV-FEL and NIR laser pulses (800 nm, $\sim$30 fs) were focused onto an effusive He gas beam introduced into a vacuum chamber through a copper tube of 1 mm inner diameter.
The XUV laser field intensity is estimated to be $\sim 1\times 10^{12}$ W/cm$^2$ from the pulse energy and the focusing geometry, with a reference to those used in the previous non-linear ionization studies on Xe and Kr \cite{Fushitani2020, Fushitani2023}.
Electron spectra were recorded using a magnetic-bottle-type spectrometer \cite{Hikosaka2010,Hishikawa2011,Matsuda2011}.
Electrons generated in the interaction region were guided into a time-of-flight (TOF) tube by inhomogeneous and homogeneous magnetic fields formed by a cone-shaped permanent magnet and a solenoid coil, respectively, and detected with a microchannel  plate (MCP) detector mounted at the end of the TOF tube.
The MCP output pulses were amplified and converted to nuclear instrumentation module (NIM) standard pulses with a timing discriminator (Ortec 9327) and recorded shot by shot using a digital oscilloscope (LeCroy WaveRunner 64Xi).

The arrival-time delay between the XUV-FEL and NIR laser pulses was measured on a shot-by-shot basis using a timing-monitor system, with a precision of 15$\pm$5 fs \cite{Owada2018,Fushitani2021}. 
Time-resolved photoelectron spectra were obtained by offline analysis sorting of single-shot photoelectron spectra according to the measured time delay. 
Electron energies were calibrated with Xe $4d$ Auger peaks \cite{Carroll2002} with an uncertainty of 0.1 eV. 
The energy resolving power was given as $R = E_e/\Delta E_e = 20$ at $E_e \sim$ 5 eV and $R = 110$ at $E_e \sim 45$ eV. 
A retardation potential, $V_\mathrm{ret}$ = 2.5 eV, was applied to the flight tube to improve the energy resolution $\Delta E_e$. 
The harmonics of FEL were suppressed by a 1.0 $\mu$m-thick Zr filter.
The XUV spectral width including the SASE fluctuation is $\Delta \omega_\mathrm{XUV}$ = 0.4 eV under the present experimental conditions.
The center photon energy, $\omega^\mathrm{center}_\mathrm{XUV}$, was set at 69.7 eV.

\section{Theoretical}
\label{sec:theory}
We solve the time-dependent Schr\"odinger equation (TDSE) for a He atom irradiated by an XUV-FEL pulse combined with a strong NIR laser field,
\begin{equation}
i\frac{\partial}{\partial t}\Psi(t)=[H_0+V(t)]\Psi(t),
\label{eq:tdse}
\end{equation}
where $H_0$ is the Hamiltonian of He atom, 
\begin{equation}
H_0=\sum_{i=1}^2 \left(-\frac{1}{2}\Delta_i+\frac{2}{r_i}\right)+
\frac{1}{|{\bf r}_1-{\bf r}_2|}
\end{equation}
and $V(t)$ is the time-dependent interaction between He and laser fields. 
We use linearly polarized fields, so the interaction within the dipole approximation in the length form can be expressed 
as 
\begin{equation}
    V(t)=(z_1+z_2) \left(f_\mathrm{NIR}(t)+f_\mathrm{XUV}(t)\right),
\end{equation}
where $f_\mathrm{NIR}(t)$ and $f_\mathrm{XUV}(t)$ are the time-dependent electric fields of
the NIR and XUV pulses, respectively. Assuming that both pulses have  
a Gaussian envelope, $f_\mathrm{NIR}(t)$ and $f_\mathrm{XUV}(t)$ can be described by the same functional form with different laser parameters indicated by different subscripts,
\begin{equation}
f(t)=F \exp{\left[-2\ln{2}\frac{t^2}{T^2}\right]} \cos{\omega t}.
\end{equation}
Here $\omega$ is the central angular frequency and $F$ is the amplitude of the electric field, and the 
peak intensity is given by $I =\frac{1}{2}\epsilon F^2$ with 
$\epsilon$ being the dielectric constant. 
The full width at half maximum (FWHM) of the temporal intensity profile is fixed at $T=413~{\rm a.u.}= 10~{\rm fs}$ for both NIR and XUV pulses.
The electric field amplitudes, $F_\mathrm{NIR}$ and $F_\mathrm{XUV}$, are fixed at 0.0045 a.u. ($2.3\times10^9$ V/m) and  0.005464 a.u. ($2.8\times10^9$ V/m), respectively. 
While the $\omega_{\mathrm{NIR}}$ is fixed at 0.05695 a.u., corresponding to a wavelength of 800 nm, we carry out calculations with different values of $\omega_\mathrm{XUV}$ for XUV, in order to take into account the photon-energy fluctuation of the SASE FEL.  

We solve Eq.(\ref{eq:tdse}) by the time-dependent hyperspherical (TDHS) method, 
which is similar to the one used for ion-He collisions \cite{Morishita2001}. 
This method contains three steps: 
(i) Setting up box-normalized basis functions in the hyperspherical coordinates. 
(ii) Solving the time-dependent equation using the box-normalized basis set and the 
calculated ground state with energy $E_0=-2.90331$ a.u. ($-79.0$ eV) as the initial state. 
(iii) Extracting the transition amplitude by projecting the wave function at the 
end of the time propagation onto the energy-normalized eigenstate of the target He atom with
the correct boundary condition. 
A more detailed description of the method has been given in Ref.~\cite{Liu2012}. 

In solving the TDSE, the wave function is expanded with respect to a basis set with $M=0$ and $S=0$, including total angular momentum up to $L=6$ and the hyperspherical channels converging up to He$^+$$(N=4)$ threshold.
We propagate the wave function from $t=-t_0$ to $t_0$, where $t_0=877.5 \,{\rm a.u.}$ (21.2 fs).
At the end of the laser pulse at $t = t_0$, we extract the transition amplitude from the time-dependent two-electron wave function
$\Psi(t_0)$ by projecting onto the energy-normalized eigenstate in the form of $\Psi_{L,j}$, where $j$ is a set of quantum numbers, $\{N,\ell,E_e,\lambda\}$. 
Here $\ell$ is the angular momentum of the bound electron,
and $\lambda$ is the angular momentum of the photoelectron. 
$E_e$ is the photoelectron energy, namely
\begin{equation}
    E_e=E_T+\frac{2}{N^2},
    \label{eq:electronenergy}
\end{equation}
where $E_T$ is the total energy. 
The transition probability to the energy-normalized eigenstate $\Psi_{L,j}$ is given by
\begin{equation}
P_{L,j}=\left|\langle \Psi_{L,j}|\Psi(t_0)\rangle\right|^2.    
\end{equation}
Then, for a given $\omega_\mathrm{XUV}$, the photoelectron spectrum with the residual ion in the state He$^+(N)$ is given as 
\begin{equation}
    S_{N}(E_e, \omega_\mathrm{XUV})=\sum_{L,\ell,\lambda}P_{L,j}.
\end{equation}

In order to compare with experimental results, theoretical spectra are convoluted to account for the spectrometer resolution and the XUV photon-energy fluctuations. 
The photoelectron spectrum convoluted with the spectrometer resolution can be expressed as
\begin{equation}
\begin{aligned}S_N'(E_e,\omega_\mathrm{XUV})&=\frac{2\sqrt{\ln{2}}}{\sqrt{\pi}w_\mathrm{S}}\int S_N(E_e',\omega_\mathrm{XUV})\\
&\times \exp{\left[-4\ln{2}\left(\frac{E_e-E_e'}{w_\mathrm{S}}\right)^2\right]}\,\mathrm{d}E_e',
\end{aligned}
\end{equation}
where $w_\mathrm{S}=(E_e-V_\mathrm{ret})/R$. 
The spectrum averaged over the XUV photon energy fluctuation is then
written as
\begin{equation}
\begin{aligned}
\bar{S}_N(E_e)&=\frac{2\sqrt{\ln{2}}}{\sqrt{\pi}w_\mathrm{X}}\int S'_N(E_e,\omega_\mathrm{XUV})\\
&\times \exp{\left[-4\ln{2}\left(\frac{\omega_\mathrm{XUV}-\omega_\mathrm{XUV}^\mathrm{center}}{w_\mathrm{X}}\right)^2\right]}\,\mathrm{d}\omega_\mathrm{XUV},
\end{aligned}
\end{equation}
where $w_\mathrm{X}=\sqrt{{\Delta\omega_\mathrm{XUV}}^2-{\delta\omega_\mathrm{XUV}}^2}$ and  $\delta\omega_\mathrm{XUV} = 4\ln{2}/T$ being the spectral FWHM of the 10-fs XUV pulse used in the calculation. 
As the center photon energy of the fluctuation, $\omega_\mathrm{XUV}^\mathrm{center} = 2.5625~\mathrm{a.u.}= 69.73~\mathrm{eV}$ is chosen to reproduce the experimental spectra.

In order to demonstrate that couplings of the $3s3p~^1P^o$ autoionizing state to  
nearby $^1D^e$ and $^1S^e$ resonances are responsible for some specific features in the 
photoelectron spectra dressed by the intense NIR laser field, we also carried out 
additional calculations with truncated bases, where the $^1D^e$ and/or $^1S^e$ quasi-bound hyperspherical channels converging to the He$^+(N \ge 3)$ thresholds are removed. 
The photoelectron spectra thus obtained  are compared with those obtained from the full basis.

\section{Results and Discussion}
\subsection{Photoelectron spectra}
Figure \ref{fig:theospectra} shows calculated photoelectron spectra of helium for decay into (a) $N$ = 2 and (b) $N$ = 1 channels at an XUV photon energy of $\omega_\mathrm{XUV}$ = 2.564~a.u. as an example. 
In the $N = 2$ spectrum, a characteristic doublet structure appears around $E_e$ = 4.4~eV.
This feature originates from the resonance excitation to the $3s3p$ $^1P^{o}$ doubly excited state from the ground state and its subsequent decay into the $N = 2$ channel. 

In the photoabsorption spectrum \cite{Domke1996}, the $3s3p$ resonance exhibits a weak asymmetric peak built on top of a continuum background.
Channel-resolved theoretical calculations \cite{Zhou1994} show that the $N = 1$ ($1s \epsilon p$) partial cross section is dominated by a large background with only a small resonance contribution, indicating that decay into $N = 1$ is largely governed by the continuum component. 
In contrast, the partial cross sections for the $N = 2$ states  ($2p \epsilon s$, $2p \epsilon d$ and $2s \epsilon p$) exhibit pronounced minima on the low-energy side of the resonance, which accounts for the asymmetric line shape in the total absorption spectrum. 

Within the perturbation theory, the XUV photoelectron spectrum to a given decay channel $N$ may be expressed as
\begin{equation}
    S_N(E_e,\omega_\mathrm{XUV}) = \sigma_N(E)S_{\mathrm{XUV}}(E, \omega_\mathrm{XUV}),
    \label{eq:spectrumshape}
\end{equation}
where $\sigma_N$ is the cross section of the photoionization to the $N$ channel given as
\begin{equation}
    \sigma_N(E)=\frac{4\pi^2\omega_\mathrm{XUV}}{c}\sum_{L=1,\ell,\lambda}{|d_{L,j}|^2},
\end{equation}
with $d_{L,j}=\langle\Psi_{L,j}|z_1+z_2|\Psi(t=-t_0)\rangle$.
The spectral profile of the XUV pulse $S_{\mathrm{XUV}}(E, \omega_{\mathrm{XUV}})$ is expressed as
\begin{equation}
\begin{aligned}
    &S_{\mathrm{XUV}}(E, \omega_\mathrm{XUV}) \\&= \frac{c\ln{2}}{2\pi\omega_\mathrm{XUV}{\delta\omega_\mathrm{XUV}}^2}  \exp\left[-4\ln{2}\left(\frac{E-\omega_\mathrm{XUV}}{{\delta\omega_\mathrm{XUV}}}\right)^2\right],    
\end{aligned}
\label{eq:xuvspectrum}
\end{equation}
where $c$ is  the speed of light and $E_e = E-I_p(N)$ with $I_p(N)=-E_0-2/N^2$ being the ionization potential of He$^+ (N)$ from the helium ground state.
The doublet structure observed in Fig.~\ref{fig:theospectra}(a) can thus be understood as a result of the interplay between the channel specific ionization minimum and the finite band width of the XUV spectrum.
As expected from the $N = 1$ partial cross section \cite{Zhou1994}, such a doublet is absent in the $N = 1$ spectrum in Fig.~\ref{fig:theospectra}(b).

When a strong NIR laser pulse is applied at a zero time-delay ($\Delta t$ = 0~fs), sideband peaks appear at energies shifted by $\omega_{\mathrm{NIR}}$ in both channels [Figs.\ref{fig:theospectra} (a) and (b)].
In addition, the $N = 2$ photoelectron spectrum exhibits two prominent resonance features near the sideband. 
Based on the energies of the doubly excited states in the $^1S^{e}$ and $^1D^{e}$ manifolds (see Table 1), these structures are assigned to resonance transitions from the $3s3p$ $^1P^{o}$ states to the `f' and `k' $^1D^{e}$ states.
This assignment is supported by the calculation where $^1D^{e}$ hyperspherical channels converging to the $N \ge 3$ thresholds are removed (``NIR w/o D").
The corresponding resonance features are strongly suppressed in the spectrum [Fig.~\ref{fig:theospectra}(d)].
Notably, these resonances to the $^1D^{e}$ states are barely visible in the $N = 1$ photoelectron spectrum in Fig.~\ref{fig:theospectra}(b), highlighting the advantage of channel-resolved study by photoelectron spectroscopy.

A closer inspection of the sideband region further reveals a small dip at $E_e \sim 6.0$~eV.
This weak feature is attributed to a resonance involving `i' $^1S^{e}$ state.
The dip disappears when $^1S^{e}$ hyperspherical channels converging to the $N \ge 3$ thresholds are removed (``NIR w/o S") [Fig.~\ref{fig:theospectra}(d)], corroborating this assignment.
While the previous theoretical study on the attosecond transient absorption spectra suggested that one-photon beatings are dominated by coupling to the $^1S^{e}$ states \cite{Petersson2017}, the resonance effects to the $^1S^{e}$ states are less prominent in the present spectra.
This difference may be attributed to the narrower band widths of the XUV and NIR laser pulses (0.4 and 0.2~eV, respectively) than those used in the previous study (5.0 and 0.68~ eV) that allows for coupling to higher resonance states in the $^1S^{e}$ manifolds.

\begin{figure*}[tb]
\includegraphics[width=17.5cm]{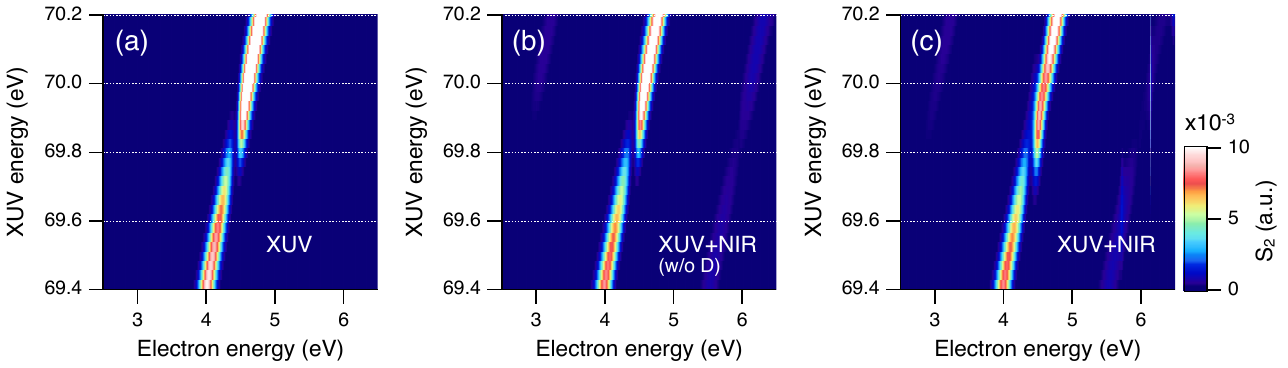}
\caption{
Theoretical photoelectron spectrum, $S_2(E_e,\omega_\mathrm{XUV})$ obtained for (a) XUV only, and for XUV+NIR calculated (b) without and (c) with the inclusion of the $^1D^e$ hypersperical channels converging to the He$^+(N\ge 3)$. 
}
\label{fig:theospectrogramraw} 
\end{figure*}

Interestingly, the NIR resonances also modify the main $3s3p$ structure [Fig.~\ref{fig:theospectra}(c)].
The theoretical spectra at different XUV photon energies are shown in Fig.~\ref{fig:theospectrogramraw}. 
The spectral minimum near 4.4~eV persists over the investigated XUV photon energy when the NIR pulse is absent [Fig.~\ref{fig:theospectrogramraw}(a)].
In contrast, a pronounced resonance associated with the `k' $^1D^{e}$ state emerges at $E_e$ = 6.14~eV for $\hbar\omega_\mathrm{XUV} \sim $70 eV and remains visible down to $\hbar\omega_\mathrm{XUV} \sim $ 69.8 eV, where the main structure is also modified as a result of the NIR dressing [Fig.~\ref{fig:theospectrogramraw}(c)].    
When the $^1D^{e}$ hyperspherical channels are removed, these dressing effects are lifted as shown in Fig.~\ref{fig:theospectrogramraw}(b).

\begin{figure}[tb]
\includegraphics[width=7cm]{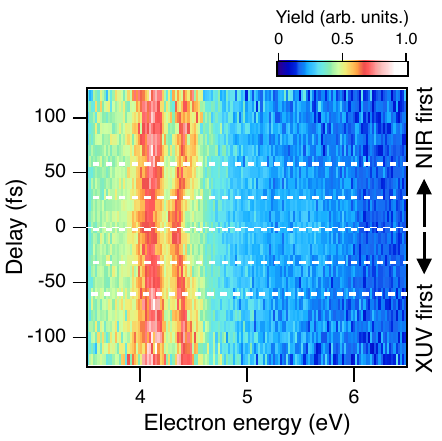}
\caption{
Experimental photoelectron spectra to He$^+$ $N = 2$, recorded as a function of the time delay $\Delta t$ between XUV (70 eV) and NIR (800 nm) laser pulses.
A clear shift of the resonance feature is visible as the temporal overlap between XUV and NIR pulses increases.
}
\label{fig:expspectrogram} 
\end{figure}

\begin{figure}[tb]
\includegraphics[width=7cm]{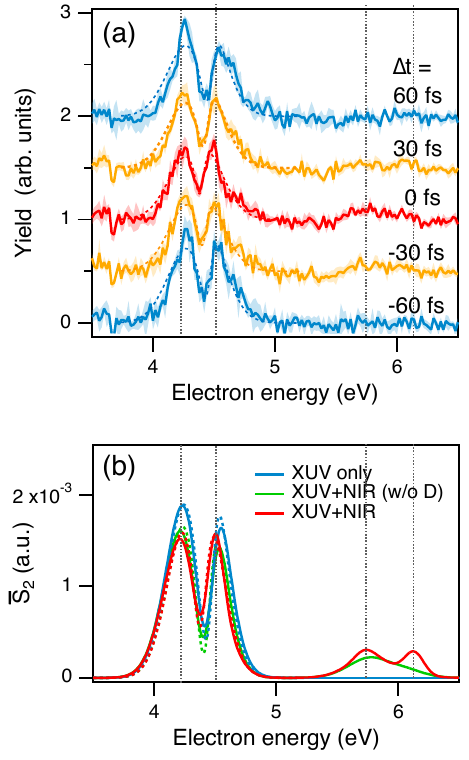}
\caption{
(a) Experimental photoelectron spectra for decay into the $N = 2$ channels of He$^+$, recorded at different pump--probe time delay $\Delta t$.
Results of the least-squares fitting to Eqs.~(\ref{eq:spectrumshape}) and (\ref{eq:multichannelFano}) also shown (dotted line).
Shaded areas indicate the statistical uncertainties estimated from six independent measurements.
(b) Theoretical $N = 2$ spectra convoluted with the SASE FEL energy bandwidth and the resolution of the photoelectron spectrometer, $\bar{S}_2(E_e)$.
Vertical lines indicate the peak energies of the theoretical spectrum obtained with XUV and NIR pulses. 
}
\label{fig:comparisonN2} 
\end{figure}

Figure \ref{fig:expspectrogram} shows the measured $N = 2$ photoelectron spectra plotted as a function of $\Delta t$.
When the XUV pulse arrives well before the NIR pulse ($\Delta t \sim -100$~fs), the spectrum minimum is located at 4.4~eV.
As the pulses start to overlap in time, the minimum shifts to lower energies and reaches its lowest value around $\Delta t$ = 0~fs.
After this, it shifts back to higher energies for positive delays.
In addition, an additional feature appears near $E_e \sim$ 5.7~eV at zero time delay. 
The spectral evolution is shown more clearly in Fig.~\ref{fig:comparisonN2}(a), where experimental photoelectron spectra are shown at selected time delays ($\Delta t = -60, -30, 0, +30$ and $+60$~fs).

These observations are consistent with the theoretical predictions in Fig.~\ref{fig:theospectra}.
For a more detailed comparison, the theoretical spectra in Fig.~\ref{fig:theospectrogramraw} are convoluted to account for the photon-energy fluctuation of SASE-FEL and with the instrumental energy resolution of the photoelectron spectrometer as detailed in Section \ref{sec:theory}. 
The obtained spectra are plotted in Fig.~\ref{fig:comparisonN2}(b) for direct comparison with the experimental spectra.
The shift of the dip energy at $\Delta t$ = 0 fs and the emergence of the resonance features near the sideband region $\sim$ 6~eV are well reproduced.

\begin{figure}[tb]
\includegraphics[width=6.8cm]{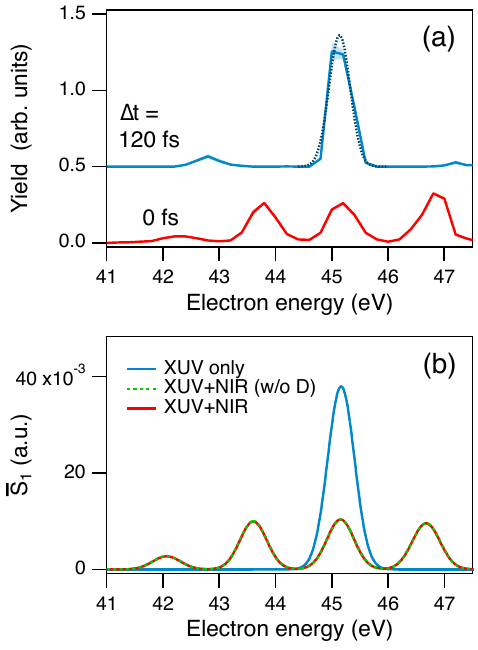}
\caption{
(a) Experimental photoelectron spectra for decay into the $N = 1$ channel of He$^+$, recorded at the pump--probe time delay $\Delta t$ = 0 and 120~fs.
Photoelectron sidebands separated by the NIR photon energy are visible at $\Delta t$ = 0~fs.
A Gaussian fit is also shown (dotted), from which the spectral resolution in this energy range is determined.  
Shaded areas indicate the statistical uncertainties estimated from six independent measurements.
(b) Theoretical $N = 1$ spectra convoluted with the SASE FEL energy bandwidth and the resolution of the photoelectron spectrometer, $\bar{S}_1(E_e)$.}
\label{fig:comparisonN1} 
\end{figure}

The corresponding $N = 1$ photoelectron spectra are shown in Fig.~\ref{fig:comparisonN1}.
First and higher-order sideband structures appear when the NIR laser pulse is present, whereas resonance features are largely absent in this channel as discussed above.
To characterize the observed spectral changes in a quantitative manner, we analyze the delay-dependent photoelectron spectra using multichannel Fano profiles. 
This approach allows us to track the evolution of the line-shape parameters, $A$ and $B$,  and the resonance energy shift $E_r$ induced by NIR dressing.

\subsection{Multichannel Fano-profile analysis}
The ${}^1P^{o}$ symmetry doubly excited states above the He$^+$ $N = 2$ threshold decay into four different continua, $1s \epsilon p$ ($N = 1$) and $2p \epsilon s$, $2p \epsilon d$ and $2s \epsilon p$ ($N = 2$). 
The NIR dressing effects observed above should therefore be treated as an intrinsically multichannel problem, in clear contrast to the previous study on doubly excited states below the $N = 2$ threshold, where only a single continuum is available.
It was shown that the partial photoabsorption cross section $\sigma_j$ can be expressed by the formula derived by Starace \cite{Starace1977} as
\begin{align}
\sigma_j(E)
& = \frac{\sigma_0^j}{1+\varepsilon^2}\{\varepsilon^2+2\varepsilon[q\mathrm{Re}(\alpha_j)-\mathrm{Im}(\alpha_j)] \notag  \\
&+[1-2q\mathrm{Im}(\alpha_j)-2\mathrm{Re}(\alpha_j)  \notag \\
&+(q^2+1)|\alpha_j|^2]\},
\end{align}
where $\sigma_0^j$ is the nonresonant background cross section and $\alpha_j$ is a complex parameter for each channel $j$.

The total cross section $\sigma_N$ for the $N= 2$ manifold is given by the sum over the three $N = 2$ channels, $j \in \{2p\epsilon s, 2p \epsilon d, 2s \epsilon p\}$,
\begin{align}
\sigma_2(E)
& = \sum_j\sigma_j,\\
&=\frac{\sigma_0}{1+\varepsilon^2}\{\varepsilon^2+A\varepsilon +B\},
\label{eq:multichannelFano}
\end{align}
where
\begin{align}
A&=2(aq-b),\\
B&=1-2bq-2a+c(q^2+1),
\end{align}
with
\begin{align}
&a=\frac{\sum_j\sigma_0^j\mathrm{Re}(\alpha_j)}{\sigma_0},\\
&b=\frac{\sum_j\sigma_0^j\mathrm{Im}(\alpha_j)}{\sigma_0},\\
&c=\frac{\sum_j\sigma_0^j|\alpha_j|^2}{\sigma_0},\\
&\sigma_0=\sum_j\sigma_0^j.
\end{align}
For the $3s3p {}^1P^{o}$ resonance, $\Gamma$ = 0.1893~eV and $q$ = 1.26 \cite{Zhou1994}.
By using the tabulated values of the complex parameters $\alpha_j$ \cite{Zhou1994}, we obtain $A$ = 1.015 and $B$ = 0.342.

By applying the expression of Eq.~(\ref{eq:spectrumshape}) with $\sigma_N$ given by Eq.(\ref{eq:multichannelFano}), we performed a least-squares fitting to the calculated $N = 2$ photoelectron spectrum $\bar{S}_2$ obtained with XUV pulse alone [Fig.~\ref{fig:comparisonN2}(b)].
Here the effective spectral width $\Delta\omega_\mathrm{XUV}$ is used in place of the XUV width $\delta\omega_\mathrm{XUV}$ in Eq.(\ref{eq:xuvspectrum}).
The profile parameters, $\sigma_0$, $A$ and $B$, and the resonance energy $E_r$ are treated as the free parameter. 
The results of the fitting are plotted in Fig.~\ref{fig:comparisonN2}(b), showing that Eq.(\ref{eq:multichannelFano}) reproduces well the spectral profile.
The extracted parameters, $A$ = 0.90(2) and $B$ = 0.36(1), are in fair agreement with the corresponding values obtained from Ref.~\cite{Zhou1994} ($A$=1.015 and $B$=0.342), given the crude approximations used in the analysis. 
The resonance energy $E_r$ = 4.47(1)~eV agrees with the tabulated value, 4.47 eV (Table \ref{tab:assignments}), demonstrating that the present procedure reliably recovers both the profile parameters and the resonance energy.

\begin{figure}[tb]
\includegraphics[width=7.5cm]{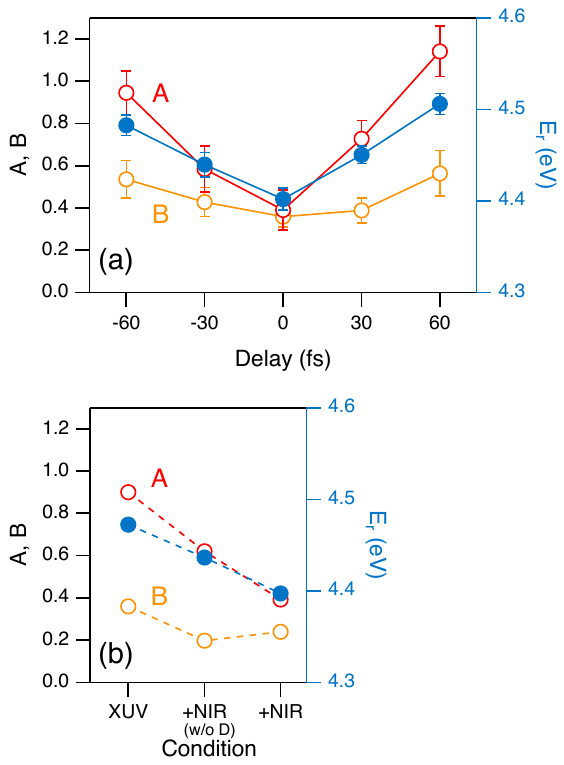}
\caption{
(a) Delay-dependent profile parameters, $A$ and $B$, and resonance energy ($E_r$) extracted by least-squares fitting of the experimental ($N = 2$) photoelectron spectra in Fig.~\ref{fig:comparisonN2}(a).
(b) Profile parameters and resonance energy obtained from the theoretical spectra in Fig.~\ref{fig:comparisonN2}(b).
}
\label{fig:abparameters} 
\end{figure}

The same analytical procedure is applied to the other theoretical spectra shown in Fig.~\ref{fig:comparisonN2}(b).
Figure \ref{fig:abparameters}(b) summarizes the determined parameters.
Both profile parameters, $A$ and $B$, change markedly when the NIR pulse is present.
With the $^1D^{e}$ states included, the extracted resonance energy decreases to $E_r \sim$ 4.40~eV.
Figure \ref{fig:abparameters}(a) shows the corresponding profile parameters and the resonance energy obtained by the least-squares fitting to the experimental spectra in Fig.~\ref{fig:comparisonN2}(a).
The profile parameters decrease as the time delay $\Delta t$ approaches zero from negative values and recover toward the original values for positive delays.
The extracted resonance energy exhibits a similar trend, reaching a minimum of $E_r$ = 4.4(1)~eV near $\Delta t \sim$ 0~fs.
The parameters at this time delay $\Delta t$ = 0~fs are reproduced by the calculations, supporting the interpretation in terms of the NIR dressing of the doubly excited states.

The NIR field leaves a clear imprint on the photoelectron spectra measured in the present work, and $E_r$ is one of the parameters that captures this effect. 
In the original formulation \cite{Starace1977}, $E_r$ represents the energy of the resonant state.
Therefore, the shift identified in our analysis suggests an NIR-induced modification of the resonance \cite{Fushitani2016}. 
In a transient absorption study of the $2p2p$ resonance \cite{Ott2014}, the ac-Stark shift in an NIR laser field was experimentally observed and was also confirmed by theoretical calculations \cite{Ott2014,Argenti2015}. 
In these studies, the shift was evaluated using a two-level model. The energy shift of an NIR-dressed state can be estimated as $\frac{\Delta E}{2}=\frac{\hbar}{2}\left(\Omega-|\Delta|\right),$where $\Omega$ is the generalized Rabi frequency, $\Omega=\sqrt{\Omega_R^2+\Delta^2}$, and $\Delta=\omega-\omega_{\mathrm{NIR}}$ is the detuning from the resonance frequency $\omega$. 
The Rabi frequency is given by $\Omega_R=\mu F$, with $\mu$ the transition dipole moment between the resonances and $F$ the NIR electric-field amplitude.

Applying this estimate to the $3s3p$ state and the `k' state in the present study yields $\Delta E/2 = 0.03~\mathrm{eV}$ for an NIR laser intensity of $0.7\times10^{12}~\mathrm{W/cm^2}$. 
Since no value has been reported for the transition dipole moment between these states, $\mu = 1$~a.u. is assumed.
The resulting $\Delta E/2$ is, at a qualitative level, comparable to that obtained from our calculations and experiments. 
The estimate should be regarded as an upper bound \cite{Ott2019}, because the corresponding Rabi periods are expected to be substantially longer than the lifetime of the $3s3p$ state (3.4~fs \cite{Rost1997}).

For the $2s2p$ resonance, theoretical calculations of the photoelectron spectra have also been reported \cite{Chu2014}.
It should be noted, however, that the Stark shift observed in attosecond transient absortion spectroscopy \cite{Ott2014} was not clearly identified in that work.
Because an absorption spectrum reflects the temporal evolution of the light-induced dipole moment, it is generally regarded as more directly sensitive to the NIR-driven atomic response \cite{Argenti2015}. 
By contrast, photoelectron spectra also contain contributions from, e.g., dressing of the continuum induced by the NIR field, which must be taken into account. 
In addition, the spectral width $\Gamma$ has been reported to vary with the NIR intensity \cite{Ivanov2003,Themelis2011}. 
Therefore, a more careful discussion is required to interpret $E_r$ as a parameter characterizing photoelectron spectra.

Beyond resonance–resonance coupling, dressing of the continua can also modify the extracted parameters.
A recent theoretical study \cite{Zielinski2015} discussed that anormalous Fano profiles with a complex $q$ parameter appear for the $2s2p$ resonance when the XUV and NIR pulses overlap.
The acceleration of electrons by the NIR laser fields redistributes amplitudes among the partial waves and modifies the Fano interference with the continuum.
The nontrivial relative phase between the bound and the continuum states would be associated with the NIR modification of the continuum state.
In our calculations, the main spectrum obtained without the $^1S^{e}$ and $^1D^{e}$ hyperspherical channels (``NIR w/o SD") still differs significantly from the XUV-only spectrum as shown in Fig.~\ref{fig:theospectra}(a), suggesting that continuum dressing contributes to the spectral modifications.
The two-photon coupling to the nonresonant ${}^1P^{o}$ multichannel continuum lying just above $N = 2$ threshold \cite{Petersson2017} may also play a role. 
These effects should be considered in interpreting the delay dependent profile parameters and resonance energy shifts. 

\section{Summary and outlook}
In summary, we investigated strong-field dressing of the doubly excited states of helium converging to the $N = 3$ threshold of He$^+$ by time-resolved photoelectron spectroscopy with synchronized XUV-FEL and strong NIR laser pulses. 
The photoelectron spectra to the $N = 2$ He$^+$ channel show clear resonance features associated with the $3s3p,{}^{1}P^{o}$ state, which exhibits a systematic delay dependence. 
Around zero-time delay ($\Delta t \sim 0$), additional resonance structures emerge in the NIR-sideband region, indicating NIR-driven coupling of the bright $3s3p$ state to nearby dark $^{1}D^{e}$ and $^{1}S^{e}$ doubly excited states. 
A multichannel Fano-profile analysis quantifies the delay-dependent evolution of the line-shape parameters and the effective resonance energy, and the observed trends are supported by theoretical calculations. 
These results establish a quantitative approach by channel-resolved photoelectron spectroscopy to characterize and control correlated two-electron resonances in strong laser fields.
This will motivate future extensions including continuum dressing and higher-order couplings for a unified description of the observed spectral modifications.

\begin{acknowledgments}
The authors thank the operation and engineering staff members of SACLA for their assistance during the beamtime. 
This work was supported by JSPS KAKENHI (Grant No JP18K03489, JP20K05549, JP21K03430,  JP24K03201) and Integrated Research Consortium on Chemical Sciences (IRCCS). 
TM was supported by JSPS KAKENHI (Grant No JP24K06916).
AH was supported by JSPS KAKENHI (Grant No JP22H00313, JP26H02267).
This research was carried out at SACLA BL1 with the approval of the Japan Synchrotron Radiation Research Institute (JASRI) (Proposal No. 2021A8052, 2022B8066, 2023A8041, 2023B8064).
\end{acknowledgments}

%

\end{document}